\documentclass [prl,twocolumn,showpacs]{revtex4}
\usepackage{dcolumn,bm}
\def\veps{\varepsilon}
\def\etal{\emph{et~al.}}
\newcommand{\eref}[1]{(\ref{#1})}
\newcommand{\Eref}[1]{Eq.~(\ref{#1})}

\begin{document}
%####################################################################
\title{High accuracy calculation of $6s \rightarrow 7s$ parity
nonconserving amplitude in Cs.}
\author{M. G. Kozlov}
\email{mgk@MF1309.spb.edu}
\author{S. G. Porsev}
\affiliation{Petersburg Nuclear Physics Institute, Gatchina, 188300, Russia}
\author{I. I. Tupitsyn}
\affiliation{Institute of Physics, St.~Petersburg State University,
Uljanovskaya 1, Petrodvorets, 198504, St.~Petersburg, Russia}
\date{\today}

\begin{abstract}
We calculated the parity nonconserving (PNC) $6s \rightarrow
7s$ amplitude in Cs. In the Dirac-Coulomb approximation our result
is in a good agreement with other calculations. Breit corrections
to the PNC amplitude and to the Stark-induced amplitude $\beta$ are
found to be $-0.4\%$ and $-1\%$ respectively. The weak charge of
$^{133}$Cs is $Q_W=-72.5 \pm 0.7$ in agreement with the standard
model.
\end{abstract}


\maketitle

%####################################################################

%###############################################
\paragraph{Introduction}
%###############################################

The last and the most accurate measurement of the ratio of the
parity nonconserving (PNC) amplitude in Cs to the Stark-induced
amplitude $\beta$ was done in Boulder~\cite{Woo97}. Later the
amplitude $\beta$ was measured with a very high
accuracy~\cite{BW99}. These measurements together with atomic
calculations of the PNC amplitude~\cite{DFS89,BSJ92} gave the most
accurate atomic test of the standard model. The authors of the the
calculations~\cite{DFS89,BSJ92} estimated their accuracy to be 1\%.
After that Bennet and Wieman~\cite{BW99} compared theoretical and
experimental results for a number of observables including
hyperfine constants, $E1$ amplitudes, and the Stark-induced amplitude
$\beta$ and suggested that the actual accuracy of atomic
calculations of the PNC amplitude was 0.4\%. Their analysis gave
the following value of the weak charge of $^{133}$Cs: $Q_W =
-72.06(44)$, which differs by $2.3\sigma$ from the standard model
prediction~\cite{PDG00}:
$$Q_W^{\rm SM} = -73.09(3)$$

This result caused an intensive discussion (see~\cite{EL00} and
references therein). It also stimulated a new wave of atomic
calculations~\cite{Der00,MGK00b,DHJS00}. First Derevianko~\cite{Der00}
suggested that previous calculations underestimated
Breit correction. He calculated the latter and added it to the PNC
amplitude from \cite{DFS89,BSJ92}. That reduced the discrepancy to
$1.0\sigma$. Somewhat smaller Breit corrections were obtained
in~\cite{MGK00b,DHJS00}.

Here we re-analyze the role of Breit interaction in calculation of
the PNC amplitude in Cs. In contrast to the paper~\cite{Der00} we
made a full scale calculation which (partly) accounts for the
higher order corrections of the many-body perturbation theory
(MBPT). The Breit interaction was included on all stages of the
calculation, i.e. on the stage of the solution of the Dirac-Fock
equations, then on the stage of the solution of the random phase
approximation (RPA) equations, and, finally, on the stage of the
evaluation of the self-energy operator. All these corrections are
of the same order of magnitude, which is not surprising taking into
account that both RPA and MBPT contributions to the answer are
quite significant. Our final value of Breit correction to the PNC
amplitude is $-0.4\%$ in agreement with our preliminary
result~\cite{MGK00b}.

We also calculated Breit correction to the Stark-induced amplitude
$\beta_{6s,7s}$ and found it to be about 1\%. That suggests that an
excellent agreement between the theory~\cite{BSJ92} and the
experiment~\cite{BW99} for this constant may be partly accidental.
In the end of this paper we discuss corrections, which are not
included in PNC calculations and conclude that the accuracy of the
atomic theory is close to 1\%. Within this accuracy there is no
contradiction with the standard model.

%################################################
\paragraph{Details of the calculation}
\label{details}
%################################################

The Dirac-Fock equations were solved on a radial grid with the help
of the code~\cite{BDT77} and all MBPT calculations were done using
summations over the basis set. We formed the Dirac-Fock-Coulomb
(DFC) and the Dirac-Fock-Coulomb-Breit (DFCB) variants of the basis
set and repeated all calculations in exactly the same way for both
of them.

It is known from previous calculations of PNC effects in Cs that
high orders in the residual Coulomb interaction are very important.
Indeed, the MBPT effects account for 13\% of the ionization potential,
for more than 60\% of the hyperfine constants, and for 22\% of the
PNC amplitude. Here we calculated explicitly only the second order
diagrams but introduced screening coefficients $C_k$ for the
Coulomb lines in these diagrams~\cite{DKPF98}. The latter can be
calculated as an average screening of the two electron Coulomb
radial integrals of a given multipolarity $k$. These coefficients
served as an approximation to the insertion of polarization
operator in Coulomb lines. The coefficients $C_k$ were chosen as
follows: $C_0=C_1=0.85,$ and $C_k=1$ for $k \ge 2$. With these
screening coefficients the Brueckner energies of the orbitals
appeared to be within 1\% from the experimental values. Still the
additional corrections were necessary in order to reach the
accuracy better than 1\%. Thus, we used the energy dependence of
the self-energy operator $\sigma(\veps)$ to improve the agreement
between calculated energy and the experiment. The optimal
one-electron energies $\veps$ appeared to be close to the Brueckner
energies of corresponding orbitals. In a similar way we chose the
energies in the RPA equations for hyperfine and PNC interactions.

The optimal choice of the one-electron energies for $\sigma(\veps)$
and for RPA matrix elements allowed us to improve the accuracy for
the lower part of the spectrum and for the hyperfine constants.
Corresponding corrections were smaller than 1\% for the energies
and 1\%~--~2\% for the hyperfine constants. RPA corrections to the
$E1$ amplitudes were much smaller and we did not do any fitting for
them. The RPA equations for weak interaction $H_P$ were solved for
the same energies as were used for hyperfine constants. That
increased PNC amplitude by 1\%.

All observables were calculated using Brueckner orbitals and RPA
amplitudes. We also calculated two smaller MBPT corrections to the
amplitudes. One of them is the so called structural radiation and
the other is the normalization correction \cite{DFS89,BSJ92}. The
former is typically of the order of few percent and the latter
reduces the final answer by approximately 1\%. There are several
corrections to the PNC amplitude of the same order of MBPT as these
ones. They correspond to the RPA-type diagrams and to the
structural radiation-type diagrams with both $E1$ and PNC vertexes
on the inner lines. These corrections were shown to be negligible
\cite{DFS89,BSJ92} and we omitted them here.

%################################################
\paragraph{Breit interaction for valence electrons}
\label{breit}
%################################################

The frequency independent Breit interaction between electrons 1 and
2 has the form
%\begin{widetext}
%-------------------------------------------------------
\begin{eqnarray}
        &&V_{\rm B} \equiv V_{\rm B}^1 + V_{\rm B}^2
        = - \frac{\bm{\alpha}_1 \cdot \bm{\alpha}_2} {r_{12}}
\nonumber \\   &&
        + \frac{1}{2} \left\{
        \frac{\bm{\alpha}_1 \cdot \bm{\alpha}_2} {r_{12}}
        -\frac{(\bm{r}_{12} \cdot \bm{\alpha}_1)
        (\bm{r}_{12} \cdot \bm{\alpha}_2)}{r_{12}^3}
        \right\},
\label{II_3}
\end{eqnarray}
%-------------------------------------------------------
where $\bm{\alpha}_i$ are Dirac matrices and $\bm{r}_{12}$ is the
distance between the electrons. The operator $V_{\rm B}^1$ is
called the Gaunt term and the operator $V_{\rm B}^2$ is called the
retardation term. For valence electrons the Gaunt term is known to
be about an order of magnitude larger then the second
one~\cite{LMY89}. The latter is of the same order as the frequency
dependent corrections to the operator~\eref{II_3} and as other QED
corrections~\cite{LTP99}. Therefore, below we neglect the second term in
\Eref{II_3}.

Calculations of Cs are done in the $V^{N-1}$ approximation and one
needs to calculate Breit interaction between the electron and the
closed shells only. It is easy to show that in this case only the
exchange term of the Breit interaction does not turn to zero. This
is almost obvious because for a closed shell
$\langle\bm{\alpha}\rangle =0$. The exchange Breit interaction can
be included in calculation either perturbatively or
self-consistently. It is known~\cite{LMY89,KPT00} that the core
relaxation results in a screening of Breit interaction between the
innermost core shells and the valence electron, significantly
reducing the final Breit corrections. In the self-consistent
approach, the potential $V_{\rm B}^{\rm core}$ should be included
in the Dirac-Fock equations. For this purpose we modified the
Dirac-Fock code~\cite{BDT77} to allow for the Coulomb-Breit
potential.

The self-consistent approach for such observables as hyperfine
constants or PNC amplitudes requires the solution of the RPA
equations. There are two types of Breit corrections to these
equations. First, one has to use DFCB orbitals and DFCB orbital
energies in RPA equations. Second, it is necessary to include Breit
interaction explicitly in two-electron matrix elements. These types
of corrections are of the same order of magnitude and should be
included on equal footing.

Finally, we calculated the the self-energy operator also using DFCB
orbitals and orbital energies. Here we did not include Breit
correction to two-electron matrix elements in diagrams for the
self-energy. These diagrams arise in the second order in residual
Coulomb interaction and always include two Coulomb lines. If in one
of these lines the Coulomb interaction is changed to the Breit one,
the resultant correction appears to be very small. The reason is
that main Coulomb corrections are associated with the interaction
of the valence electrons with the upper core shells, while the most
important Breit interaction is with the innermost core shells.

All calculations with the exception for structural radiation and
normalization corrections were repeated two times in exactly the
same manner in Dirac-Coulomb and Dirac-Coulomb-Breit
approximations. That allowed us to single out Breit corrections
even when they were comparable to the errors caused by the
incompleteness of the basis set or by the approximations we used to
account for higher order terms of MBPT.

%################################################
\paragraph{Numerical results}
\label{results}
%################################################

The PNC amplitude of the $6s \rightarrow 7s$ transition can be
written as
% a sum over intermediate states:
\begin{widetext}
\begin{equation}
  E1_{\rm PNC}(6s,7s) =
  \sum_n{\left(
  \frac{\langle 7s_j,m|E1|np_j,m\rangle
        \langle np_j,m|H_P|6s_j,m\rangle}
  {\veps_{6s_j}-\veps_{np_j}}
  +
  \frac{\langle 7s_j,m|H_P|np_j,m\rangle
        \langle np_j,m|E1|6s_j,m\rangle}
  {\veps_{7s_j}-\veps_{np_j}}
  \right)},
\label{III_0}
\end{equation}
\end{widetext}
where $j=1/2$. This amplitude is sensitive to the energy spectrum
and $E1$-amplitudes as well as to the matrix elements of the weak
interaction $H_P$. The latter depends on the wave function in the
vicinity of the nucleus and in this respect is similar to the
matrix elements of the hyperfine interaction. Thus, in order to
estimate the accuracy of the calculation of the PNC amplitude one
has to analyze the accuracy for all these observables.

%####################################################################
\begin{table}
\caption{Binding energies for the lower levels of Cs (au). The columns
DF and BA correspond to the Dirac-Fock and Brueckner
approximations. The column OE corresponds to Brueckner
approximation with self-energy calculated at `optimal' energies
$\veps(6s)=-0.22$, $\veps(7s)=-0.15$, and $\veps(np)=-0.10$.
Finally we give Breit corrections to the energy in Brueckner
approximation. }

\label{tab_e}

%\begin{ruledtabular}
\begin{tabular}{lrrrrr}
\hline
\hline
&\multicolumn{1}{c}{DF}
&\multicolumn{1}{c}{BA}
&\multicolumn{1}{c}{OE}
&\multicolumn{1}{c}{Breit}
&\multicolumn{1}{r}{Exper.}   \\
\hline
$6s_{1/2}$ & 0.12737  & 0.14424  & 0.14312  &$ 0.00002$  & 0.14310 \\
$6p_{1/2}$ & 0.08562  & 0.09226  & 0.09217  &$-0.00004$  & 0.09217 \\
$6p_{3/2}$ & 0.08379  & 0.08967  & 0.08959  &$-0.00001$  & 0.08965 \\
$7s_{1/2}$ & 0.05519  & 0.05869  & 0.05864  &$ 0.00000$  & 0.05865 \\
$7p_{1/2}$ & 0.04202  & 0.04389  & 0.04386  &$-0.00001$  & 0.04393 \\
$7p_{3/2}$ & 0.04137  & 0.04305  & 0.04303  &$ 0.00000$  & 0.04310 \\
\hline
\hline
\end{tabular}
%\end{ruledtabular}
\end{table}
%####################################################################
%####################################################################
\begin{table}
\caption{Hyperfine constants for $^{133}$Cs (MHz).}

\label{tab_h}

%\begin{ruledtabular}
\begin{tabular}{ldddddd}
\hline
\hline
& \multicolumn{1}{c}{DF}
& \multicolumn{1}{c}{MBPT}
& \multicolumn{1}{c}{OE}
& \multicolumn{1}{c}{Breit}
& \multicolumn{1}{l}{Theory}
& \multicolumn{1}{l}{Exper.}  \\
\hline
$6s_{1/2}$ & 1424  & 2350  & -52   & +5.0  &  2302   & 2298   \\
$6p_{1/2}$ & 160.9 & 298.6 & -5.3  & -0.2  &   293.5 & 292    \\
$6p_{3/2}$ &  23.9 &  53.4 & -2.3  & -0.0  &    51.2 & 50.3   \\
$7s_{1/2}$ & 391.4 & 546.8 & -0.8  & +0.8  &   546.8 & 546    \\
$7p_{1/2}$ &  57.6 & 94.2  & -0.2  & -0.0  & 94.0    & 94.3   \\
$7p_{3/2}$ &   8.6 & 17.1  & -0.0  & -0.0  & 17.1    & -      \\
\hline
\hline

\end{tabular}
%\end{ruledtabular}
\end{table}
%####################################################################

Tables~\ref{tab_e} and~\ref{tab_h} present our results for the
energies and hyperfine constants of the lower levels of Cs. These
tables show results on the different stages of calculation
including the initial Dirac-Fock approximation, the MBPT
and Breit corrections. We also give corrections associated with
the optimization of the energies in MBPT and RPA calculations
as described above.

Breit corrections to $E1$ amplitudes appear to be mostly smaller than
0.1\%. Similar corrections to scalar polarizabilities of the levels
$6s$ and $7s$ are about $-0.1$\%. These polarizabilities can be
used to calculate the Stark shift $\delta \nu_{6s,7s}$ for the $6s
\rightarrow 7s$ transition (in Hz/(V/cm)$^2$):
%-------------------------------------------------------
\begin{eqnarray}
        \delta \nu_{6s,7s} &=&
        \left\{\begin{array}{ll}
        .7268 & \mbox{without Breit,} \\
        .7259 & \mbox{with Breit,} \\
        .7223 & \mbox{theory \cite{BSJ92},} \\
        .7262 & \mbox{experiment \cite{BRW99}.}
        \end{array} \right.
\label{III_1}
\end{eqnarray}
%-------------------------------------------------------

Breit correction to the Stark-induced vector amplitude
$\beta_{6s,7s}$ is more important because this amplitude turns to
zero in the non-relativistic approximation. Numerical values for
this amplitude in au are:
%-------------------------------------------------------
\begin{eqnarray}
        \beta_{6s,7s} &=&
        \left\{\begin{array}{ll}
        27.17 & \mbox{without Breit,} \\
        26.89 & \mbox{with Breit,} \\
        27.0(2)  & \mbox{theory \cite{BSJ92},} \\
        27.02(8) & \mbox{experiment \cite{BRW99}.}\\
        \end{array} \right.
\label{III_1a}
\end{eqnarray}
%-------------------------------------------------------
In the experiment~\cite{BRW99} the amplitude $\beta$ was measured
relative to the hyperfine $M1$ amplitude. Dzuba and Flambaum
suggested~\cite{DF00} that the latter is slightly smaller than was
assumed in~\cite{BRW99} and gave the value $\beta=26.96(5)$.

Our results for the PNC amplitude are presented in
table~\ref{tab_pnc}.
%####################################################################
\begin{table}
\caption{PNC amplitude
$E1_{\rm PNC}(6s,7s)$ in the units $i\cdot 10^{-11}
Q_W/(-N)$~au.% I and II correspond to two terms in \Eref{III_0}.
}

\label{tab_pnc}

%\begin{ruledtabular}
\begin{tabular}{lcccccc}
\hline
\hline
&&\multicolumn{1}{c}{~~DF}
& \multicolumn{1}{c}{~MBPT}
& \multicolumn{1}{c}{~~OE}
& \multicolumn{1}{r}{Breit}
& \multicolumn{1}{r}{Total} \\
\hline        %   DF        MBPT        OE        Breit     Total
This work   & & $-.742$  & $-.896$  & $-.905$  &   .004   &$-.901$ \\
Dzuba \etal   &
\cite{DFS89}  &          & & &  \multicolumn{1}{c}{---}   &$-.908$ \\
Blundell \etal&
\cite{BSJ92}  &        &\multicolumn{2}{d}{-.907}& .002   &$-.905$ \\
Derevianko    &
\cite{Der00}  &          &          &          &   .008   &        \\
Dzuba \etal   &
\cite{DHJS00} & $-.739$&\multicolumn{2}{d}{-.907}& .005   &$-.902$ \\
\hline
\hline
\end{tabular}
%\end{ruledtabular}
\end{table}
%####################################################################
Calculations were done for the uniformly charged nucleus of the
radius 6.20~fm. Following~\cite{JS99,Der00} we introduced $-0.1$\%
correction to account for the difference in neutron and charge
radii of the nucleus.

We see that {\it ab initio} calculation and calculation with
`optimal' energies for the self-energy operator and for RPA
amplitudes differs by 1\%, i.e. PNC amplitude appears to be less
sensitive to the variation of the energies used for the self-energy
operator $\sigma(\veps)$ and for RPA equations than the hyperfine
constants. Thus, we assume that uncertainty for the PNC amplitude
associated with the fitting is smaller than for the hyperfine
constants.

%################################################
\paragraph{Discussion}
\label{discussion}
%################################################

For the Dirac-Coulomb approximation our value for the PNC amplitude
is in a good agreement with the results of the
calculations~\cite{DFS89,BSJ92}. Our value of the Breit correction
is approximately two times smaller than the result of Derevianko.
This discrepancy is almost eliminated if we add $-0.003$ correction to
his result, which is caused by the Breit contribution to the energies
of atomic levels. Derevianko argues that this correction should not be
included if one fits the energies semiempirically.

We first want to note that calculations~\cite{DFS89,BSJ92} did not
use any energy fit. Therefore, Breit correction to these calculations
should definitely include this contribution.
In our calculation we \textit{did} use the energy fit and thus
Derevianko's argument seems to work here. However, we did not
insert experimental energies into \Eref{III_0}. Instead, we
modified the MBPT by changing the argument of the self-energy
operator $\sigma(\veps)$. That changed the energies \textit{and}
the Brueckner orbitals thus affecting the numerators as well as
denominators in \Eref{III_0}. We think that this is a more
consistent way to account for the high orders of the MBPT. We
repeated the energy fit after Breit interaction was included. That
changed Breit corrections to different contributions to the PNC
amplitude, but the overall correction still remained $-0.4$\%.

The analysis of the numerical data from
Tables~\ref{tab_e},~\ref{tab_h} and Eqs.~\eref{III_1}
and~\eref{III_1a} shows that the energies, the Stark-shift
$\delta\nu$, and even the Stark-induced amplitude $\beta$ are in a
very good agreement with the experiment. However, the accuracy for
the hyperfine constants of the levels $6s$ and $6p_{1/2}$ is
significantly lower. In order to improve these constants we had to
make the energy fit which changed both constants by approximately
2\%. The reason for the low accuracy here is obvious: the MBPT
corrections to $A(6s)$ and $A(6p_{1/2})$ are 65\% and 81\%
correspondingly. Taking into account that MBPT correction to the
PNC amplitude is only 22\% and that the energy fit changes this
amplitude only by 1\%, we estimate our accuracy for the PNC
amplitude to be about 0.5\%.

This estimate does not include uncertainties associated with the
nuclear structure and QED corrections. QED corrections to the
hyperfine structure and the uncertainties associated with the
nuclear magnetic moment distribution were discussed
in~\cite{MP95,Shab99,Sush00}. The authors conclude that there are
several corrections on the scale of 0.1--0.3\%.

For the PNC amplitude the vacuum polarization correction was
estimated in~\cite{LS94,BLPSK99} to be 0.2--0.4\% for $Z=55$. The
core relaxation can screen it significantly, as it does with the
Breit correction, so it is rather an order of magnitude estimate of
the QED corrections to the PNC amplitude. Another uncertainty of the
order of 0.1--0.2\% is caused by the poor knowledge of the neutron
distribution in the nucleus~\cite{JS99,Der00}.

In our analysis we use Gaunt approximation to Breit operator
while in the paper~\cite{DHJS00} the retardation term was also
included.
%The matrix elements of the omitted retardation term are an order of
%magnitude smaller, but they may become important because of the
%significant screening of Breit interaction by the core and by
%cancellations of different contributions to the PNC amplitude.
This may explain the difference between our value 0.004 and the value
0.005 obtained in~\cite{DHJS00}. As we mentioned above, the retardation
term is of the same order of magnitude as the frequency-dependent
corrections to the Breit operator and as other QED corrections.
Therefore, the difference between our results can serve as an estimate
of the accuracy of Breit approximation. Thus, the total uncertainty for
the PNC amplitude is about 1\% and there is no meaningful difference
with the standard model. Our final value for the weak charge is:  $$
Q_W = -72.5(7)$$

\begin{acknowledgments}
We are grateful to Derevianko, Dzuba, Labzowsky, Mosyagin, Shabaev,
Titov, and Trzhaskovskaya for helpful discussions. This work was
partly supported by RFBR grants No 98-02-17663 and No 00-03-33041.
\end{acknowledgments}

%#####################################################
%\bibliographystyle{apsrev}
%\bibliography{pnc,my_ref_w}
%#####################################################

%#####################################################
\end{document}